# Room-temperature antiferromagnetic CrSe monolayer with tunable metal-insulator transition in ferroelectric heterostructures

Xiao-Sheng Ni,[1,#] Yue-Yu Zhang,[2,#] Dao-Xin Yao,[1] and Yusheng Hou [1]

**AFFILIATIONS**

[1] Guangdong Provincial Key Laboratory of Magnetoelectric Physics and Devices, State Key Laboratory of Optoelectronic Materials and Technologies, Center for Neutron Science and Technology, School of Physics, Sun Yat-Sen University, Guangzhou, 510275, China

[2] Wenzhou Institute, University of Chinese Academy of Sciences, Wenzhou, Zhejiang 325001, China; Oujiang Laboratory (Zhejiang Lab for Regenerative Medicine, Vision and Brain Health), Wenzhou, Zhejiang 325001, China

**ABSTRACT**

Recently, there has been a rapidly growing interest in two-dimensional (2D) transition metal chalcogenide monolayers (MLs) due to their unique magnetic and electronic properties. By using an evolutionary algorithm and first-principles calculations, we report the discovery of a previously unexplored, chemically, energetically, and thermodynamically stable 2D antiferromagnetic (AFM) CrSe ML with a Néel temperature higher than room temperature. Remarkably, we predict an electric field-controllable metal-insulator transition (MIT) in a van der Waals (vdW) heterostructure comprised of CrSe ML and ferroelectric $Sc_2CO_2$. This tunable transition in $CrSe/Sc_2CO_2$ heterostructure is attributed to the change in the band alignment between CrSe and $Sc_2CO_2$ caused by the ferroelectric polarization reversal in $Sc_2CO_2$. Our findings suggest that 2D AFM CrSe ML has important potential applications in AFM spintronics, particularly in the gate voltage conducting channel.

[#] These authors contributed equally to this work.

Authors to whom correspondence should be addressed: houysh@mail.sysu.edu.cn and yaodaox@mail.sysu.edu.cn

Transition metal chalcogenides (TMCs) have garnered significant attention due to their diverse range of properties, including magnetism, superconductivity, and optoelectronics [1-4]. These compounds have the chemical formula $M_mX_n$, where M and X represent a transition metal element (e.g., Mo and W) and a chalcogen element (e.g., S, Se and Te) [5], respectively. As the dimensionality of TMCs decreases from bulk to monolayer (ML), many physical phenomena can emerge. For instance, when $MoS_2$, one of the most well-known TMCs, is thinned from bulk to ML, an indirect to direct bandgap transition is observed [6]. Moreover, when the transition metal M in TMCs is magnetic, exciting physical properties can be designed in their MLs [7,8]. Recently, Aapro *et al*. synthesized MnSe ML, which is an antiferromagnetic (AFM) material with singular characteristics that make it suitable for spintronics applications [9]. Similarly, Cr, another transition metal with a large magnetic moment, has demonstrated outstanding performance in its 2D compounds, such as the emerging 2D ferromagnetism in $CrI_3$ ML [10,11]. To explore 2D TMCs with Cr, Zhang *et al*. synthesized a quasi-2D ferromagnetic (FM) CrSe film with a thickness of approximately 2.5 nm and a Curie temperature of 280 K [12]. Intriguingly, first-principles calculations reveal that this layered structure can be transformed from a FM metal to a room-temperature half-metal via chemical edge modification [13]. Besides, a pentagonal AFM CrSe ML is proposed by utilizing atomic transmutation from pentagonal building blocks [14]. However, it can be argued that realistic 2D TMC MLs with Cr cannot be synthesized using conventional methods like bulk cleaving or atomic transmutation since crystal structures are highly likely to change when the dimensionality is reduced. Therefore, more advanced material design methods are of importance to explore 2D TMC MLs with Cr beyond conventional practices.

The ability to control the electronic properties of 2D TMC MLs externally is crucial for enriching their technological applications in ultra-thin electronic devices. Previous studies have shown that strain and external electric fields can effectively control the electronic properties of TMCs, and even lead to a metal-insulator transition (MIT) [15-18]. However, these theoretically predicted MITs in TMCs require a large external strain up to 10% or a very high electric field of 10 V/nm [15,19,20], which is challenging to achieve in practice. On the other hand, 2D TMC MLs exhibit various emerging properties when they form

Page **2** of **13**

heterostructures with substrates [21-23]. Therefore, building heterostructures of 2D TMC MLs with other functional substrates, such as 2D ferroelectric materials, could provide exciting opportunities to engineer the electronic properties of 2D TMC MLs for spintronics applications [23-26].

In this study, we combine a multi-objective global optimization algorithm and first-principles calculations to perform structural searches on 2D TMC MLs composed of Cr and Se. We discover that the ground-state structure of CrSe ML has a space group (SG) *P4/nmm*. Its dynamical stability is confirmed by the absence of imaginary frequency in its phonon spectrum. Our systematic density functional theory (DFT) calculations and Monte Carlo (MC) simulations demonstrate that CrSe ML is a high Néel temperature ($T_N$ = 522 K) antiferromagnet with a narrow band gap. We predict a MIT in van der Waals (vdW) heterostructure of CrSe ML and ferroelectric $Sc_2CO_2$ when the ferroelectric polarization of $Sc_2CO_2$ is reversed. By analyzing the band alignments of $CrSe/Sc_2CO_2$ with opposite ferroelectric polarizations, we reveal that this MIT is caused by the band alignment change from the type-III to the type-I. Our study introduces a magnetic member into the 2D TMCs family and provides insight for designing innovative 2D switching spintronic devices.

Taking inspirations from the experimentally synthesized MnSe ML [9] and the magnetism observed in 2D materials containing Cr [27,28], we carry out a structural search on the TMC ML composed of Cr and Se elements with a 1:1 chemical ratio. The chemical stability of CrSe ML is ensured by the composition of $Cr^{2+}$ cations and $Se^{2-}$ anions. In addition to the *P-3m1* phase as reported for MnSe ML, our global optimization package IM$^2$ODE identifies eight unreported structures with different SGs (see the inserts in Fig. 1). These structures are potential candidates in the structural phase diagram of CrSe ML (structural information can be found in Part IV in the supplementary material). Figure 1. shows the total energies of all the structures discovered in this study and those reported in literatures, including the *P3m1* phase in Refs. [13,29], the *P4/m* phase in Ref. [14] and the *P-3m1* phase derived from the structure of MnSe ML using the conventional atomic transmutation method. From Fig. 1, it can be observed that the ground-state structure of CrSe ML is a tetragonal *P4/nmm* phase. Importantly, this *P4/nmm* phase has a much lower total energy than the *P-3m1* phase which is a MnSe ML transmutation. It is worth noting that the *P3m1* phase reported in Ref. [13] is actually a metastable phase.

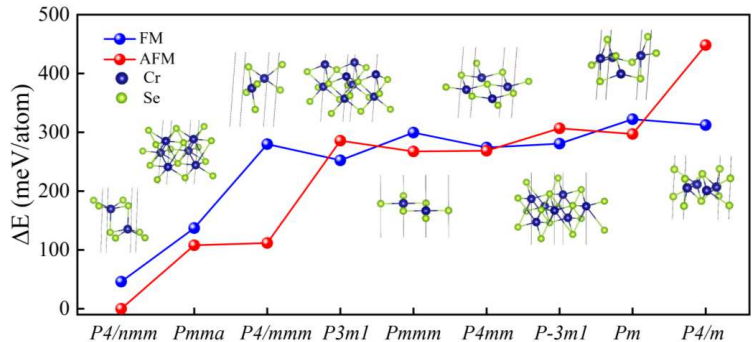

*FIG. 1. Structural phase diagram of CrSe ML. The energy differences between different structures are obtained by taking the P4/nmm phase with an AFM order as reference. The insets show the crystal structures of CrSe ML with different SGs. Blue and green spheres represent Cr and Se atoms, respectively.*

Here, we investigate the thermodynamic stability of the *P4/nmm* phase which is the ground-state structure of CrSe ML. Firstly, we analyze its crystal structure in detail. Unlike MnSe ML, which has a honeycomb lattice, CrSe ML features a square lattice in the *ab* plane (Fig. 2a-2b). As depicted in Fig. 2a, two layers of $Cr^{2+}$ cations are situated between two layers of $Se^{2-}$ anions which are at the top and bottom surfaces. Each $Cr^{2+}$ cation is coordinated by five $Se^{2-}$ anions, and vice versa. The phonon spectrum indicates that the *P4/nmm* phase is dynamically stable, because no imaginary frequencies are present (Fig. 2c). Additionally, the crystal structure of CrSe ML is thermally stable, as evidenced by the AIMD simulated structures and total energies as a function of time at 300 K (Fig. S3 in supplementary material). Therefore, the thermodynamical stability of the *P4/nmm* phase of CrSe ML is confirmed by both phonon spectrum and AIMD simulations. This indicates that the *P4/nmm* phase of CrSe ML can be synthesized in future experiments. Henceforth, we will focus on the *P4/nmm* phase of CrSe ML unless otherwise stated.

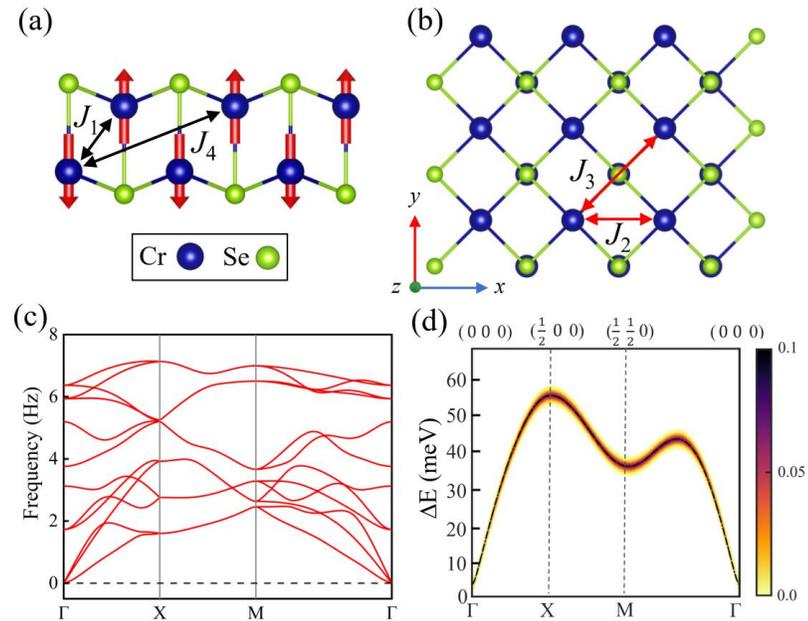

*FIG. 2. Crystal structure, phonon spectrum and magnetic properties of the P4/nmm phase of CrSe ML. (a) Side and (b) top views of the crystal structure. Blue and green spheres represent Cr and Se atoms, respectively. The red arrows depict the magnetic moments of $Cr^{2+}$ cations. The exchange paths of the Heisenberg exchange parameters $J_i$ (i=1,2,3, and 4) are shown by double-arrowed lines in (a) and (b). (c) and (d) show the phonon and spin-wave spectra of CrSe ML, respectively.*

After determining the crystal structure of CrSe ML, we proceed to investigate its magnetic properties. DFT calculated magnetic moments are 3.8 $\mu_B$/Cr, indicating that magnetic $Cr^{2+}$ cations possess a spin S = 2. To determine the magnetic ground state, we employ a spin Hamiltonian that consists of Heisenberg exchange interactions and magnetic anisotropy energy (MAE). Explicitly, the spin Hamiltonian takes the form of

$$H_{\text{spin}} = \sum_{ij} J_{ij} S_i \cdot S_j - K \sum_i \left(S_i^z\right)^2 \qquad (1).$$

In Eq. (1), $J_{ij}$ is the Heisenberg exchange parameters between spins $S_i$ and $S_j$, and $K$ is the MAE constant. Here, a negative (positive) $J_{ij}$ means a FM (AFM) interaction and MAE originates from the single-ion anisotropy. We employ a least-squares fit technique [30] to calculate the Heisenberg exchange parameters. Our calculations show that the nearest neighbor (NN) $J_1$ and fourth-NN $J_4$ are 3.14 and 1.57 meV, respectively. This indicates that the interactions between $Cr^{2+}$ cations in different layers are AFM. Moreover, second-NN $J_2$ and third-NN $J_3$ are -1.43 and -3.21 meV, respectively, suggesting FM interactions between $Cr^{2+}$ cations in the same layer. Finally, the calculated MAE constant $K$ is 0.12 meV, which implies an out-of-plane magnetic easy axis in CrSe ML. To summarize, the magnetic interactions of CrSe ML are dominated by Heisenberg exchange interactions and a non-negligible single-ion anisotropy.

Based on the aforementioned magnetic interaction parameters, we conduct Monte Carlo (MC) simulations [31,32] to determine the magnetic ground state of CrSe ML. Surprisingly, we find that the magnetic ground state is an unusual AFM order that consists of two antiferromagnetically coupled FM planes, as shown in Fig. 2a. This AFM order is the same as that in MnSe ML [9], despite the fact that CrSe and MnSe MLs have different in-plane lattices. Furthermore, the MC-simulated Néel temperature is as high as 522 K, suggesting that CrSe ML is a possible room-temperature 2D antiferromagnet. To gain more insights into the magnetic properties of CrSe ML, we study its spin-wave spectrum using the linear spin wave approximation. As illustrated in Fig. 2d, the spin wave energy can reach up to approximately 55 meV at the X point, indicating the presence of strong Heisenberg exchange interactions in CrSe ML. Notably, the non-negligible single-ion anisotropy leads to a spin-wave gap of approximately 4 meV. This large spin-wave gap is crucial to counteracting thermal fluctuations and highly beneficial to realizing a high Néel temperature in CrSe ML.

Figure 3a presents the band structure and density of states (DOS) of the AFM CrSe ML when spin-orbit coupling (SOC) is not included in DFT calculations. Due to the AFM order, spin-up and spin-down bands are completely degenerate. The conduction and valence bands are cone-shaped and somewhat flat, respectively. At the Γ point, they nearly touch with a small direct band gap of 25 meV. The DOS shown in Fig. 3a indicates that the states from $Cr^{2+}$ cations contribute significantly to the bands near the Fermi level. According to the point group $D_{4h}$ of SG *P4/nmm*, the five *3d* orbitals of $Cr^{2+}$ cations are split into $A_{1g}$ ($d_{3y^2-r^2}$), $B_{1g}$ ($d_{x^2-y^2}$), and $B_{2g}$ ($d_{xy}$) singlet states, as well as $E_g$ ($d_{xz}$, $d_{yz}$) doublet states. From the projected DOS of $Cr^{2+}$ cations (Fig. 3a), the $B_{2g}$ ($d_{xy}$) singlet state has the highest energy. According to Hund's rule, the $A_{1g}$, $B_{1g}$, and $E_g$ states are fully occupied by the four *3d* electrons of $Cr^{2+}$ cations, producing a magnetic moment of 4 $\mu_B$/Cr. This is consistent with DFT calculated results (3.8 $\mu_B$/Cr). Upon including SOC in DFT calculations, we find that the band gap of CrSe ML is enlarged to 98 meV. It is worth noting that the profiles of the band structures with and without SOC are similar (Fig. 3a-3b). Therefore, SOC plays a critical role in the semiconducting characteristics of CrSe ML.

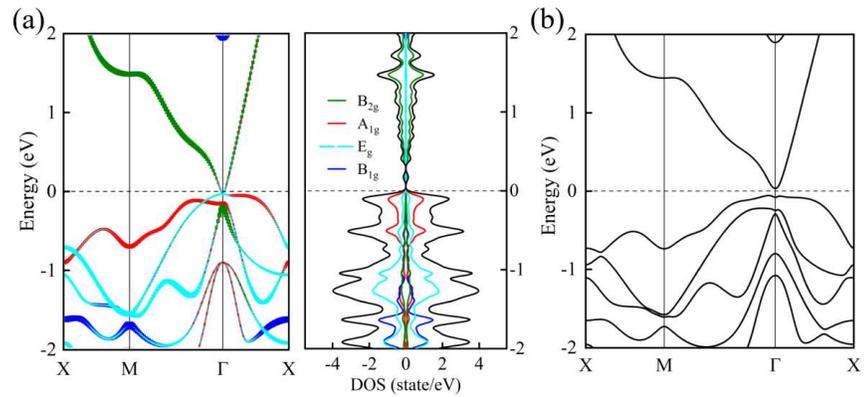

*FIG. 3. (a) Band structure and DOS of the AFM CrSe ML when SOC is not included in DFT calculations. Bands and DOS are projected to $B_{1g}$ (blue), $E_g$ (cyan), $A_{1g}$ (red) and $B_{2g}$ (dark green) states. The widths of lines in the band structure indicate the weights of the different states. (b) Band structure of the AFM CrSe ML when SOC is considered in DFT calculations. The Fermi level is set to zero and indicated by the dashed black lines in (a) and (b).*

To design intriguing properties in CrSe ML, we construct its vdW heterostructure with $Sc_2CO_2$ ML. We choose $Sc_2CO_2$ ML due to its atomically thin ferroelectricity and large out-of-plane polarization, which can reach up to 1.60 $\mu C/cm^2$ [33,34]. To overcome the lattice mismatch between CrSe and $Sc_2CO_2$ MLs, we employ a 5 × 1 supercell for the former and a $3\sqrt{2} \times \sqrt{2}$ supercell for the latter. While keeping the lattice constant of the former unchanged, we adjust the lattice constant of the later to match the former. This results in compressive strains of 0.5% and 3.5% on *a* and *b* axes of $Sc_2CO_2$, respectively. In $CrSe/Sc_2CO_2$, CrSe ML can pair with either the upward-polarization (P↑) or downward-polarization (P↓) state of $Sc_2CO_2$ ML. We consider several highly symmetric stacking configurations (see Part VI in supplementary material) to obtain the energetically favorable ones for $CrSe/Sc_2CO_2$ with P↑ (CrSe/P↑) and P↓ (CrSe/P↓) states. In the most stable configuration of CrSe/P↑, some Se atoms in the bottom layer locate on the top of Sc atoms, while some Cr atoms in the same layer sit on the top of O and Sc atoms (Fig. 4a). In the most stable configuration of CrSe/P↓, some Se atoms in the bottom layer locate on the top of C atoms, while some Cr atoms in the same layer sit on the top of C and Sc atoms in CrSe/P↓ (Fig. 4c). Hereafter, we discuss only the magnetic and electronic properties of the most stable configurations for both CrSe/P↑ and CrSe/P↓.

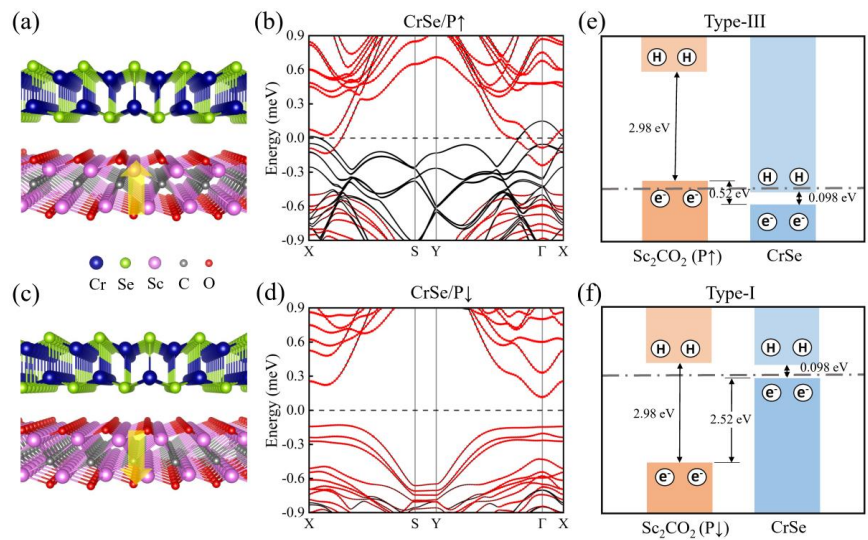

FIG. 4. (a) Side view of CrSe/P↑. (b) Band structure of CrSe/P↑ calculated by DFT with SOC. The red dots indicate the states from CrSe. (c) and (d) same as (a) and (b) but for CrSe/P↓. The upward and downward yellow arrows in (a) and (c) denote the ferroelectric polarization of $Sc_2CO_2$. In (a) and (c), Cr, Se, Sc, C and O atoms are represented by blue,

*green, purple, gray and red balls, respectively. (e) and (f) show the schematic representations of the band alignments of CrSe/P↑ and CrSe/P↓ when SOC is included in DFT calculations, respectively.*

By comparing the total energies of different magnetic orders, we obtain that the magnetic ground states of both CrSe/P↑ and CrSe/P↓ are the AFM orders with two ferromagnetically coupled planes (see Part VII in supplementary material). Note that this AFM orders are identical to those of the free-standing CrSe ML. Hence, the magnetic ground state of CrSe ML is not changed by the presence of the ferroelectric $Sc_2CO_2$ ML in CrSe/$Sc_2CO_2$. Figure 4b and 4d show the band structures of CrSe/P↑ and CrSe/P↓, respectively. It is evident that the Fermi level intersects bands from both CrSe and $Sc_2CO_2$ MLs in CrSe/P↑ (Fig. 4b), indicating that CrSe/P↑ is a metal. Conversely, there is a distinct band gap in the band structure of CrSe/P↓ (Fig. 4d). Notably, the band gap of CrSe/P↓ (250 meV) is much larger than that of the free-standing CrSe ML (98 meV). The band gap enlargement of CrSe/P↓ is attributed to structural deformation in CrSe which is caused by the presence of $Sc_2CO_2$, because the lattice constant of the CrSe ML is kept unchanged in building CrSe/P↓ (see Part VIII in supplementary material). Overall, we predict a MIT in CrSe/$Sc_2CO_2$ when the polarization of the ferroelectric $Sc_2CO_2$ is reversed from upward to downward. This finding stands in contrast to the improved optical absorptions [35,36], molecular magnetism control [34], and ferroelectric control of electronic structures [37,38] as reported in previous studies of semiconductor and $Sc_2CO_2$ heterostructures. In practice, the MIT in CrSe/$Sc_2CO_2$ can be easily modulated by voltage [19], providing potential applications in valve devices [19,39].

As the characteristic of band alignment is crucial to determining the trend of charge transfer in heterostructures, we study the band alignment in CrSe/$Sc_2CO_2$ to make clear the physical mechanism behind the MIT. The core-level alignment method [40] is employed to obtain the band alignment. Figure 4e and 4f show the relative positions of the band edges with respect to the core energies in CrSe/P↑ and CrSe/P↓, respectively. The valence band offset in CrSe/P↑ (Fig. 4e) crosses the Fermi level, resulting in a charge transfer and leading to the metallic nature. In contrast, the Fermi level lies between the conduction bands of CrSe

Page **9** of **13**

ML and the valence bands of $Sc_2CO_2$ ML in CrSe/P↓ (Fig. 4f), thus prohibiting any charge transfer. Overall, CrSe/P↑ has a zero-gap type-III band alignment and is metallic, but CrSe/P↓ has a type-I band alignment and is a semiconductor. Therefore, the MIT in CrSe/$Sc_2CO_2$ arises from the change in its band alignment from zero-gap type-III to type-I when the ferroelectric polarization of $Sc_2CO_2$ is reversed from upward to downward.

From an experimental standpoint, it is feasible to fabricate the *P4/nmm* phase of CrSe ML and its vdW heterostructure with $Sc_2CO_2$. Firstly, our structural search establishes the *P-3m1* phase as the ground-state structure of MnSe ML by means of our global optimization package IM$^2$ODE (Fig. S1 in supplementary material). Note that this ground-state structure is consistent with recent experimental findings [9]. Thus, the *P4/nmm* phase of CrSe ML found by the same package can be synthesized experimentally in the future. On the other hand, $Sc_2CO_2$ can be synthesized by fully oxidizing 2D MXene $Sc_2C$ [34,41]. Particularly, 2D MXene $Ti_2CO_2$, a sibling of $Sc_2CO_2$, has been fabricated [42]. Hence, the experimental synthesis of $Sc_2CO_2$ is also feasible. After the synthesis of both CrSe and $Sc_2CO_2$ MLs, they can be mechanically stacked to form vdW heterostructure CrSe/$Sc_2CO_2$ due to their vdW nature [43].

In summary, our study utilizes a combination of global optimization search and DFT calculations to determine that the ground-state structure of 2D CrSe ML has space group *P4/nmm*. We also demonstrate its thermodynamic stability through phonon spectrum and *ab initio* molecular dynamics simulations. The magnetic ground state of this unreported CrSe ML is an AFM order consisting of two antiferromagnetically coupled FM planes and it has a high Néel temperature of 522 K. Interestingly, the band alignment between CrSe and $Sc_2CO_2$ MLs in vdW heterostructure CrSe/$Sc_2CO_2$ changes from type-III to type-I when the ferroelectric polarization of $Sc_2CO_2$ is reversed from upward to downward. This change leads to a controllable MIT in CrSe/$Sc_2CO_2$, which can be modulated through the application of an external electric field. Our findings suggest that the room-temperature 2D antiferromagnet CrSe ML is experimentally feasible and its heterostructure with $Sc_2CO_2$ could have promising applications in valve devices.

See supplementary material for the details of computational methods, the results of structural search on MnSe ML, the effect of different $U_{eff}$ on the structural ground state of CrSe ML, structural information of CrSe ML with different SGs, *ab initio* molecular dynamics simulations of the *P4/nmm* phase of CrSe ML, different stacking configurations, magnetic orders of CrSe/Sc$_2$CO$_2$ heterostructures and comparisons between the free-standing CrSe ML and the CrSe ML directly isolated from CrSe/P↓.

We thank Trinanjan Datta for helpful discussions. This project is supported by NKRDPC-2022YFA1403301, NSFC-12104518, NKRDPC-2022YFA1402802, GBABRF-202201011118, NSFC-92165204, NSFC-11974432, GBABRF-2022A1515012643, GBABRF-2019A1515011337, Shenzhen International Quantum Academy (Grant No. SIQA202102), Leading Talent Program of Guangdong Special Projects (201626003), the Startup Grant of Wenzhou Institute and Oujiang Laboratory (WIUCASQD2021014 and WIUCASQD2022025) and the Open Project of Guangdong Provincial Key Laboratory of Magnetoelectric Physics and Devices (No. 2022B1212010008). Calculations are performed at Tianhe-II.

## AUTHOR DECLARATIONS

**Conflict of Interest**

The author has no conflict of interest to declare.

**Authors Contributions**

**Xiao-Sheng Ni:** Investigation (equal); Methodology (equal); Writing – original draft (equal). **Yue-Yu Zhang:** Investigation (equal); Methodology (equal); **Dao-Xin Yao:** Conceptualization (equal); Investigation (equal); Funding acquisition (equal); Supervision (equal); Writing – review and editing (equal). **Yusheng Hou:** Conceptualization (equal); Funding acquisition (equal); Investigation (equal); Project administration (equal); Resources (equal); Supervision (equal); Writing – review and editing (equal).

## DATA AVAILABILITY

The data that support the findings of this study are available from the corresponding author upon reasonable request.